\documentclass[prl,aps,twocolumn,floats,superscriptaddress]{revtex4}
\usepackage{amsmath}
\usepackage{graphicx}
\usepackage{epsfig}

\def \s{\sigma}     
   
   \def \d{\delta}

\def \Lam{\Lambda}

\def \ord{\mathcal{O}}

\def\dg{^\dagger}
\def\lba{\left(}    
\def\rba{\right)}
\def\lbc{\left[}
\def\rbc{\right]}

\def\ord{\mathcal{O}}

\def\be{\begin{equation}}
\def\ee{\end{equation}}
\newcommand \bea {\begin{eqnarray} }
\newcommand \eea {\end{eqnarray}}
\newcommand \sign {\hbox{sign}}

\begin{document}

\title{Regular and Singular Fermi Liquid Fixed Points
    in Quantum Impurity Models}

\author{Pankaj Mehta}
\affiliation{Center for Materials Theory, Rutgers University, Piscataway, NJ 08855, U.S.A. }  
\author{L. Borda}
\affiliation{Sektion Physik and Center for Nanoscience, LMU M\"unchen,
Theresienstrasse 37, 80333 M\"unchen, Germany}
\affiliation{Hungarian Academy of Sciences, Institute of Physics,
TU Budapest, H-1521, Hungary}

\author{Gergely Zarand}
\affiliation{Theoretical Physics Department, Budapest University of Technology and Economics,
Budafoki ut 8. H-1521 Hungary }

\author{Natan Andrei}

\author{P. Coleman}
\affiliation{Center for Materials Theory, Rutgers University, Piscataway, NJ 08855, U.S.A. }

\begin{abstract}

We show that thermodynamics is insufficient to probe the nature of the
low energy dynamics of quantum impurity models and a more subtle
analysis based on scattering theory is required.  Traditionally,
quantum impurity models are classified into one of two categories:
Fermi liquids and non-Fermi liquids, depending on the analytic
properties of the various thermodynamic quntities. We show, however,
that even when a quantum impurity model is a Fermi liquid (an incoming
electron at the Fermi level scatters elastically off the impurity),
one may find singular thermodynamic behavior if characteristics of
quasiparticles are {\it not analytic} near the Fermi surface.
Prompted by this observation, we divide Fermi liquids into two
categories: regular Fermi liquids and singular Fermi Liquids. The
difference between regular Fermi liquids, singular Fermi liquids , and
non-Fermi liquids fixed points is explained using properties of the
many-body $S$-matrix for impurity quasiparticle scattering.  Using the
Bethe-Ansatz and numerical RG, we show that whereas the ordinary Kondo
Model is a regular Fermi liquid the underscreened Kondo model is a a
singular Fermi liquid. This results in a breakdown of Nozi\`eres'
Fermi liquid picture for the underscreened and explains the singular
thermodynamic behavior noticed in Bethe Ansatz and large-$N$
calculations.  Furthermore, we show that conventional regular Fermi
liquid behavior is re-established in an external magnetic field $H$,
but with a density of states which diverges as $1/H$.  Possible
connections with the field-tuned quantum criticality recently observed
in heavy electron materials are also discussed.

\end{abstract}

\maketitle

\section{Introduction}

Quantum impurity models have been studied extensively in condensed
matter physics both experimentally and theoretically.  Many approaches
have been developed to characterize their low energy
physics. Conventionally, systems are classified into one of two
categories, Fermi liquid (FL)\cite{noz} and non-Fermi Liquid (NFL)
depending on their low temperature specific heat behavior.  In
particular, systems with non-integer power dependence on temperature
are called NFLs. In this paper, we show that thermodynamics is
insufficient to probe the nature of the low energy dynamics and a more
subtle analysis based on electron-impurity scattering theory is
required.  We illustrate our ideas using recent as well as established
results on the underscreened Kondo Model (UKM).

In a Fermi liquid, the low energy dynamics are 
described in terms of well defined quasi-particles close to a Fermi-surface. 
Furthermore, for an impurity problem, the quasiparticles at the Fermi energy 
are isomorphic to the original electron states. For this reason,
electrons at the Fermi  energy scatter {\em
elastically} off the impurity: both the ingoing and outgoing states
consist of a single electron.  However, in a generic non-Fermi liquid
impurity system, this is not the case. Even when the incoming
electrons are on the Fermi surface they can scatter inelastically; an
incoming electron state {\em does not} scatter into a single outgoing
electron state, but instead, excites a large variety of collective modes
including particle-hole excitations. For example, in the extreme case
of the  two-channel Kondo model, the out going scattering state does not 
include any single electron component after 
scattering with the impurity \cite{Ludwig}.

This difference between Fermi liquids and non-Fermi liquids manifests
itself most clearly in the renormalization group (RG)
flow of the single particle matrix
 elements of the many-body $S$-matrix, $\langle k\;\mu, {\rm in}|\hat
 S |k'\;\mu', {\rm in}\rangle$ where $k$ and $k'$ denote the momenta
 of incoming and outgoing electrons and $\mu, \mu'$ the rest of their
 quantum numbers. In the space of single electrons states it is easy
 to show (see below) that the matrix elements depend only on $\omega =
 k = k'$ where $\omega$ denotes the energy of incoming particle
 measured with respect the the Fermi energy. Unitarity requires
 $S(\omega)$ to be a complex number with modulus less than equal to one.
 Using RG  one obtains the behavior of the theory at $\omega=0$. 
For a Fermi liquid $|S(\omega=0)|=1$ implying that at the Fermi level,
the inelastic scattering cross section vanishes  and
single particle at the  scattering is completely
characterized by a phase shift. On the other
hand, for a non-Fermi liquid model,  $|S(\omega=0)|<1$. Consequently,
NFLs have a non-vanishing many particle scattering rate and a finite
inelastic scattering cross section at the Fermi surface.

\begin{widetext}

\begin{figure}[t]
\includegraphics[width=0.6\columnwidth, clip]{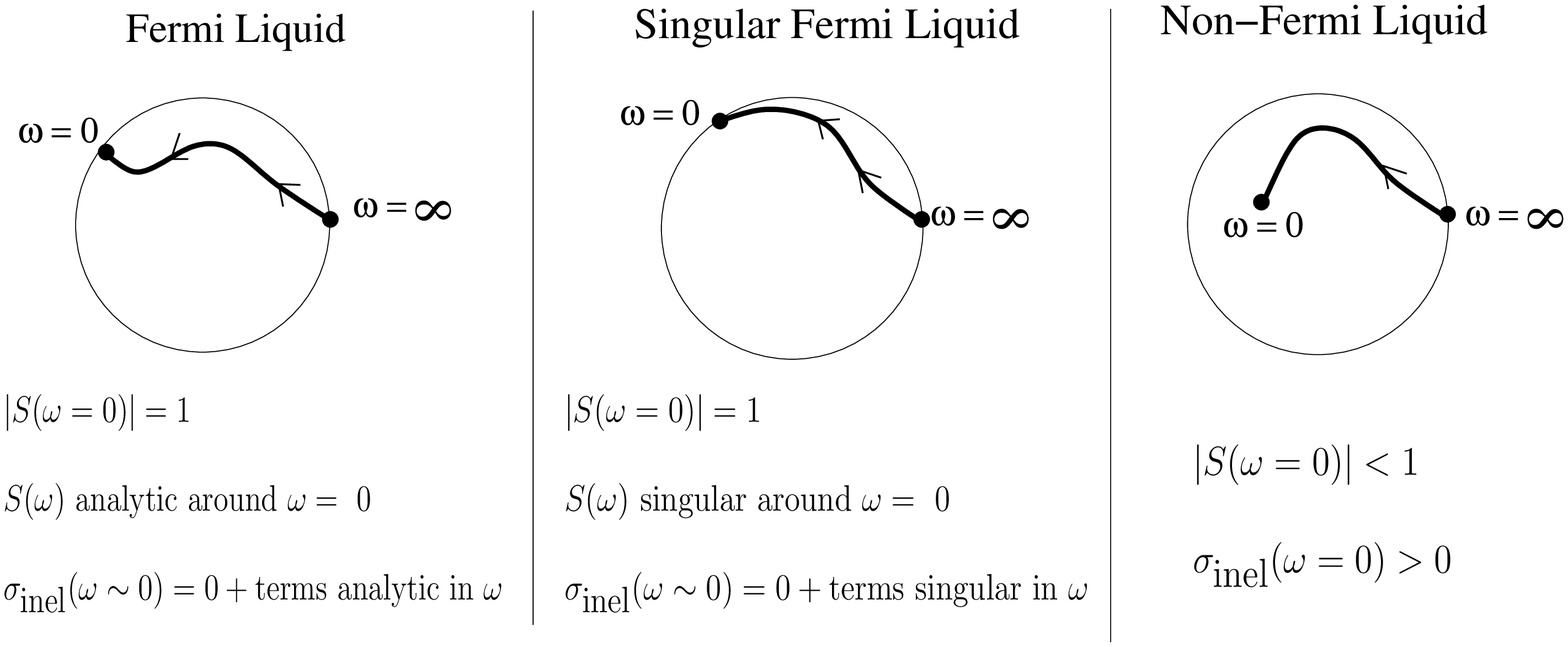}
\caption{
\label{fig:singular}
Sketch of the renormalization group flows of the eigenvalues of the single particle $S$-matrix.  The eigenvalues are within the unit circle.
Particles at high enough energies ($\omega \to \infty$)  do not see the impurity, therefore $S\to 0$ in this limit. Inelastic scattering 
processes are allowed whenever $|S(\omega)| < 1$.  In Fermi liquids  at the Fermi energy $|S(\omega=0)| = 1$, implying the absence of inelastic 
scattering of electrons. 
For non-Fermi liquids $|S(\omega=0)| < 1$, while for singular Fermi liquids
 $S(\omega)$ approaches the unit circle non-analytically as $\omega \to 0$.
}
\end{figure}

\end{widetext}

We shall argue below that even when a quantum impurity is a Fermi-liquid 
in the sense described above one may still find singular 
thermodynamic behavior.  This
occurs when characteristics of quasiparticle are {\it not
analytic} near the Fermi surface. In terms of scattering, this means
that the eigenvalues of the S-matrix for impurity quasi-particle
scattering approach the unit circle non-analytically as the quasi-particle
energy $\omega$ approaches the Fermi level.  
For this reason, it is necessary to divide Fermi liquids into two types: 
{\em regular Fermi liquids} (RFL) and {\em singular Fermi liquids} (SFL).  
In the former, the eigenvalues of the single particle S-matrix approach
the unit circle  analytically whereas in the latter, they approach it
singularly. For both types of fixed points, the single particle $S$-matrix 
is unitary and 
an incoming electron at the Fermi surface scatter elastically off the
impurity. However, the two types of FLs exhibit very different
phenomenological properties. A regular Fermi liquid exhibits the usual
properties associated with FLs, and hence, by an abuse of notation, we
shall often omit the term `regular' when referring to this class of
fixed points.  On the other hand, SFLs exhibit a wide variety of
behavior not ordinarily associated with Fermi liquids such as extreme
sensitivity to applied fields and a divergent specific heat.This classification scheme and
the main properties of the three impurity classes are summarized in
Fig.~\ref{fig:singular}.

An example of the difference between regular and singular FLs is seen in
the striking difference in the strong coupling physics of the ordinary
Kondo Model (KM) and the underscreened Kondo Model (UKM) (see
~\cite{varkondo} and references within). The UKM describes the interaction
of a magnetic impurity with spin $S>1/2$ with a sea of conduction
electrons. In the UKM, the impurity and the electrons are coupled by
anti-ferromagnetic interaction. At low temperatures, the impurity spin
is therefore partially screened from spin S to spin $S^*=S-1/2$. What
distinguishes the UKM from the ordinary Kondo model is the residual
magnetic moment that remains even after screening. This residual
moment couples ferromagnetically to the remaining conduction
electrons. Though the ferromagnetic coupling is irrelevant, it tends
to zero very slowly.  As a result, there is a subtle interplay between
the residual moment and the electron fluid that leads to radically
different physics from the ordinary KM at the strong-coupling fixed
point of the UKM.

A Bethe-Ansatz and a large $N$ analysis of the underscreened Kondo
model revealed that at zero field, this system exhibits singular
behavior, with a divergent specific heat coefficient $C_v/T$ at zero
field \cite{sacramento,CPe}. In a finite field, the linear specific
heat coefficient was found to diverge as $1/(B \; {\rm ln}^2
(T_K/B)$. To study the scattering properties of the model we
re-examine it using the Bethe-Ansatz and the numerical renormalization
group (NRG). From the Bethe-Ansatz solution, we find that the
sacttering matrix and density of states of spinons at the impurity
(DOS) is a singular function of quasi-particle energy in zero magnetic
field. The DOS becomes analytic in finite fields; however, it shows a
singular behavior as the magnetic field scales to zero.  These results
are confirmed using the numerical renormalization group (NRG)
calculations, where we can directly compute the phase shift of spin
1/2 electron excitations from the finite size spectrum. The singular
nature of the DOS results in the breakdown of Nozi\`eres' picture of
the strong coupling fixed point and indicates that the physics of the
UKM and the ordinary Kondo model are quite different.  However,
despite this singular behavior,the NRG calculations confirm that the
fixed point finite size spectrum of the UKM is that of a Fermi liquid,
{\em i.e.}, it can be characterized by a simple phase shift $\pi/2$.

The analysis we present here may also be relevant to heavy fermion
systems. Over the years, much of new insight obtained
in heavy electron physics has been acquired  from the 
study of simplified, impurity models ~\cite{4}.
Anderson's original model for the formation of local moments, is itself
an impurity model. Doniach's Kondo lattice scenario for 
heavy  electron metals was based on an understanding of the
impurity Kondo model, long before approximate solutions to the lattice
were available ~\cite{doniach}.   The idea to use a large $N$ expansion
for the Kondo lattice model, grew from a corresponding study of single
impurity models ~\cite{largeN} and early motivation for the understanding of 
non-Fermi liquid behavior  in Uranium heavy fermion systems grew from an application of the
two-channel Kondo model to these systems ~\cite{cox}. Most recently,
impurity models have played  a role in proposed models for the quantum critical behavior of 
heavy electron systems ~\cite{qcp} .

In recent experimental studies, heavy 
electron materials were fine-tuned away from  an antiferromagnetic quantum critical point (QCP) using 
a magnetic field\cite{ gegenwart, Custers},  
revealed that parameters of the heavy Fermi liquid can be field-tuned. In particular,  
the temperature dependent properties of the system near the QCP were shown to depend only on the 
ratio $T/(B-B_c)$. Therefore one can draw parallels with the field tuned change in  behavior of the 
UKM from a SFL to a regular FL.

This paper is structured as follows. In Section~2, we discuss the general 
classification of regular and singular Fermi Liquids and the application of this 
classification scheme for Kondo models in more detail. In section~3,  we use 
the Bethe-Ansatz to calculate the DOS and find that it
is singular in the absence of a magnetic field.
In Section~4, we present Numerical Renormalization Group calculations 
confirming our Bethe-Ansatz results. In Section~5, we discuss the breakdown
of  Nozi\`eres Fermi liquid picture for the UKM. Finally,
we discuss connections with field-tuned criticality 
and some future lines of research. Some details of the Bethe Ansatz 
calculation are given in Appendix~I.

\section{Singular Fermi Liquids and Non-Fermi Liquids}

In a Fermi liquid impurity model the quasiparticle excitations at the
Fermi level become identical to undressed conduction electrons. In
other words, an incoming electron at the Fermi energy scatters into a
single outgoing electron.  In contrast, in a non-Fermi liquid impurity
model, such as the two-channel Kondo model, quasiparticle excitations
do exist, but are orthogonal to the original incoming electrons. This
implies that in a non-Fermi liquid impurity model only a fraction of
an incoming electron scatters into a single electron state, with the
rest exciting particle-hole excitations via inelastic scattering.  In
the extreme case of the two-channel Kondo model, {\em e.g.}, the
outgoing state can be shown to be completely orthogonal to a single
electron state.\cite{Maldacena,JanZar}

These properties can be most easily captured through the many body
$S$-matrix, which we shall analyze in the rest of this section. The
discussion of the $S$-matrix will allow us, in particular, to
distinguish between non-Fermi liquid and singular Fermi liquid
systems.

Let us consider a general quantum impurity problem at $T=0$
temperature, described by the following Hamiltonian
\begin{equation}
H = -i \sum_\mu \int dx\; :\psi_\mu^{\dagger}(x,t) \partial_x \psi_\mu(x,t):
+ H_{\rm int}\;.
\label{eq:H_total}
\end{equation}
Here the fields $\psi_\mu$ are chiral one-dimensional fermions, and
usually represent radial excitations in some three-dimensional angular
momentum channel coupled to the impurity. The label $\mu$ represents
those discrete internal degrees of freedom (spin, flavor, crystal
field, angular momentum indices, etc.)  that may couple to the
impurity. The precise form of the impurity-fermion interaction,
$H_{\rm int}$ is of no importance for the purpose of our discussion
below.

The central quantity  we are interested in is the many-body  $S$-matrix, 
$\hat S$,  defined in terms of incoming and outgoing scattering states, 
$|a\rangle_{\rm in}$ and $| b\rangle_{\rm out}$ as (see e.g. \cite{Itzikson}) 
\begin{equation} 
\langle b, {\rm out} |\;a, {\rm in}\rangle \equiv \; 
\langle b, {\rm in}|\hat S|\;a, {\rm in}\rangle\;.
\end{equation}
The 'in' and 'out' states are eigenstates of the total Hamiltonian,
Eq.~(\ref{eq:H_total}), satisfying the boundary conditions that they
tend to plane waves in the $t\to -\infty$ and $t\to \infty$ limits,
respectively. In the interaction representation, the explicit form of
the $S$-matrix is given by the well-known expression
$\hat S =  {\rm T}\exp \{-i \int_{-\infty}^\infty  H_{\rm int} (t) \;dt\}$,
where ${\rm T}$ is the time ordering operator, and the interaction
$H_{\rm int} (t)$ is adiabatically turned on and off during the time
evolution.  In the following, we shall only be interested in the
single particle matrix elements of the $S$-matrix,
\begin{equation}
\langle k\;\mu, {\rm in}|\hat S |k'\;\mu', {\rm in}\rangle
= 2\pi \;\delta(k-k')\; {\cal S}_{\mu \mu'}(\omega)\;,
\label{eq:onshellS}
\end{equation}
where $k$ and $k'$ denote the momenta of incoming and outgoing
 electrons.  In Eq.~(\ref{eq:onshellS}) we separated a Dirac delta
 contribution due to energy conservation and thus defined the on shell
 single particle $S$-matrix, $\cal S(\omega)_{\mu \mu'}$, with $\omega
 = k$ the energy of the incoming and outgoing
 electrons  \cite{Itzikson}.

Unitarity of the $S$-matrix poses severe constraints on the eigenvalues 
$s_\lambda(\omega) \equiv
r_\lambda(\omega) e^{i 2 \delta_\lambda(\omega)}$
of $\cal S$, which must be within the unit circle: 
\begin{equation}
|s_\lambda(\omega)| = r_\lambda(\omega)  \le 1\;.
\end{equation}

If $\cal S$ has an eigenvalue that is {\em not} on the unit circle, 
this implies that one can construct an incoming  single particle state 
which with some probability scatters {\em inelastically} into a multi-particle outgoing 
state. 
To show this more explicitly, let us consider the $T$-matrix defined through 
\[
\hat S = \hat 1 + i \;\hat T\;.
\]
We can then define the on-shell $T$-matrix ${\cal T}(\omega)_{\mu\mu'}$ analogous to 
Eq.~(\ref{eq:onshellS}), and the corresponding eigenvalues  are simply given by
\begin{equation}
\tau_\lambda(\omega) = -i \;(s_\lambda(\omega)-1)\;.
\end{equation}  
As discussed in Ref.~\onlinecite{dephasing}, 
the knowledge of the single particle matrix elements of the many-body $T$-matrix enables 
us to compute the total scattering cross section off the impurity 
in the original three-dimensional impurity problem through the optical theorem as
\begin{equation}
\sigma_{\rm tot} = \sigma_0    \sum_\lambda 2 \;|\varphi_\lambda(\omega)|^2  
{\rm Im}\{ \tau_\lambda(\omega)\}\;,
\end{equation}
where $\sigma_0 = \pi/k_F^2$  with $k_F$ the Fermi momentum, and 
$\varphi_\lambda(\omega)$ denotes the wave function amplitude of the incoming electron in scattering 
channel $\lambda$.   
{\em Elastic scattering} off the impurity can be defined as single particle 
scattering processes where the outgoing state consists of a single outgoing electron.  
The elastic scattering cross section is simply proportional 
to the  square of the elements of the $T$-matrix, and is given by 
\begin{equation}
\sigma_{\rm el} = \sigma_0  \sum_\lambda  |\varphi_\lambda(\omega)|^2 |\tau_\lambda(\omega)|^2\;.
\end{equation}
Having determined both $\sigma_{\rm tot}$  and $\sigma_{\rm el}$, we can
define the {\em inelastic} scattering cross section off the impurity 
as the difference of these cross sections,\cite{JanZawa} 
$\sigma_{\rm inel} = \sigma_{\rm tot} - \sigma_{\rm el}$,
\begin{equation}
\sigma_{\rm inel} 
= \sigma_0  \sum_\lambda  |\varphi_\lambda(\omega)|^2  (1- r_\lambda(\omega)^2)\;.
\label{eq:inel}
\end{equation}

We  define a quantum impurity model to be of {\em non-Fermi liquid} type 
if some of the eigenvalues of the single particle $S$-matrix are not on the 
unit circle in the $\omega\to 0 $ limit. By Eq.~(\ref{eq:inel}) this immediately implies that non-Fermi liquid models have the unusual property that  
even electrons at the Fermi energy can scatter off the impurity 
inelastically  with a finite probability. 

Typical examples of non-Fermi liquid models are given by over-screened multichannel 
Kondo models. In the two channel Kondo model, {\em e.g.}, it has been shown 
in Refs.~\cite{Maldacena} and ~\cite{JanZar} using bosonization methods that  
the single particle matrix elements of the S-matrix identically vanish at the Fermi energy, 
immediately implying that $r_\lambda =0$  and thus $\sigma_{\rm el} =
\sigma_{\rm inel} = \sigma_{\rm tot}/2$  at the Fermi energy \cite{footnote,dephasing}. 

Most other models, however, such as screened or under-screened Kondo
models, the Anderson impurity model, the resonant level model, or
over-screened models in an external field, fall in the category of
{\em Fermi liquids}, since in all these cases all eigenvalues of the
single particle $S$-matrix are located {\em on the unit circle} at
$\omega=0$.  This implies that in Fermi liquids electrons at the Fermi
energy scatter completely {\em elastically} off the impurity, and that
this scattering can be characterized in terms of simple phase shifts.

The structure of the energy dependence of the $s_\lambda(\omega)$'s,
{\em i.e.} the renormalization group flow of the eigenvalues of the
single particle $S$-matrix, however, does depend on the specific Fermi
liquid model, and allow for further classification: We can define as
{\em singular Fermi liquids} those models, where the convergence to
the $\omega=0$ Fermi liquid fixed point is singular in $\omega$, while
we shall call {\em regular Fermi liquids} those, where the
convergence is analytical.  By these terms, the standard spin 1/2
Kondo model is a regular Fermi liquid, while under-screened Kondo
models such as the $S=1$ single channel Kondo model studied in this
paper belong to the class of {\em singular Fermi liquids}. We shall
see below that singular Fermi liquids have singularities in the low
energy thermodynamic properties while having only elastic (albeit
singular elastic) scattering on the Fermi surface.

The properties of the flows of the eigenvalues of the single particle  
scattering matrix $\cal S$ have been  summarized for the  three classes 
of quantum impurity models  in Fig.~\ref{fig:singular}. 

External perturbations such as a magnetic field, {\em e.g}, 
can usually generate a cross-over to a regular Fermi liquid state in non-Fermi liquid or 
singular Fermi liquid models. However, the parameters of the resulting 
regular Fermi liquid may depend non-analytically on the external 
perturbations, and the corresponding Fermi liquid energy scale 
vanishes in the absence of them. Therefore the properties of these Fermi liquids become singular as a function 
of the external perturbation. The $S>1/2$ single channel 
Kondo model studied here, {\em e.g.},  becomes a regular Fermi liquid in a magnetic field, however, 
the phase shifts $\delta_s(\omega = 0)$ exhibit logarithmic singularity as a function of 
the  magnetic field, corresponding to a divergent impurity DOS in the  $H\to0$ limit.

\section{Bethe Ansatz Calculation of the Density of States for the Underscreened Kondo Model}

We proceed to show that the UKM is an example of a singular FL by
analyzing the density of states, DOS, and the  phase shift of electrons 
scattering off the impurity. We show that in zero magnetic field
the DOS is a singular function of quasiparticle
energy. 
We also show that while $|s(\omega)| \to 1$ as $\omega \to 0$ the limit is
approached in a singular manner. Finally, we study the effect of a
magnetic field and show that the singularity in the DOS is cut off by
a finite field.

The Hamiltonian for the UKM can be mapped to the following
one-dimensional Hamiltonian

\begin{equation}
H_{UKM} = -i \sum_{a}\int dx \, \psi_{a}^\dag(x) 
\partial_x \psi_{a}(x) 
+ J\lba \psi_{a}^{\dag}(0) \vec{\sigma_{ab}} \psi_b(0)\rba \vec{S} \;, \nonumber\end{equation}
where $\psi_a\dg(x)$ is the creation operator of an electron with spin
$a$, and $\vec S$ is a localized spin at the origin coupled to the
electron sea by an antiferromagnetic coupling $J$. In this equation
the left-moving chiral Fermions $\psi_a\dg(x)$ in regions $x>0$ and
$x<0$ simply represent the incoming and outgoing parts of the
conduction electrons' $s$ wave function in the three-dimensional
problem.

The spectrum of the UKM can be determined from Bethe-Ansatz solution
\cite{under, wiegmann, natan}. The excitations consist of uncharged  spin-1/2
excitations, spinons, and spinless charge excitations, holons. In the
spinon-holon basis, the wavefunction for the electron can be written
as a sum of products of a spin wavefunction and a charge wavefunction. Since
the Kondo interaction affects only the spin sector, we will ignore the
charge sector in the analysis that follows.

In the spin sector, an electron can be expressed as a {\em
superposition} of spinons and anti-spinons. Formally, this is done
through a form-factor expansion of the electron onto the spinon
basis. At low-energies, the coefficients of the the multi-spinon terms
in the form factor expansion tend to zero.  For this reason, at
sufficiently low energies, it is a reasonable to approximate the
electron by a spinon \cite{natan82}. Since we are interested in the 
low-energy properties of the UKM, we will employ this approximation.  The
validity of this approximation will be checked by comparing our
results at the Fermi energy with those of the Numerical
Renormalization Group (NRG).

\begin{figure}[t]
\includegraphics[width=0.6\columnwidth]{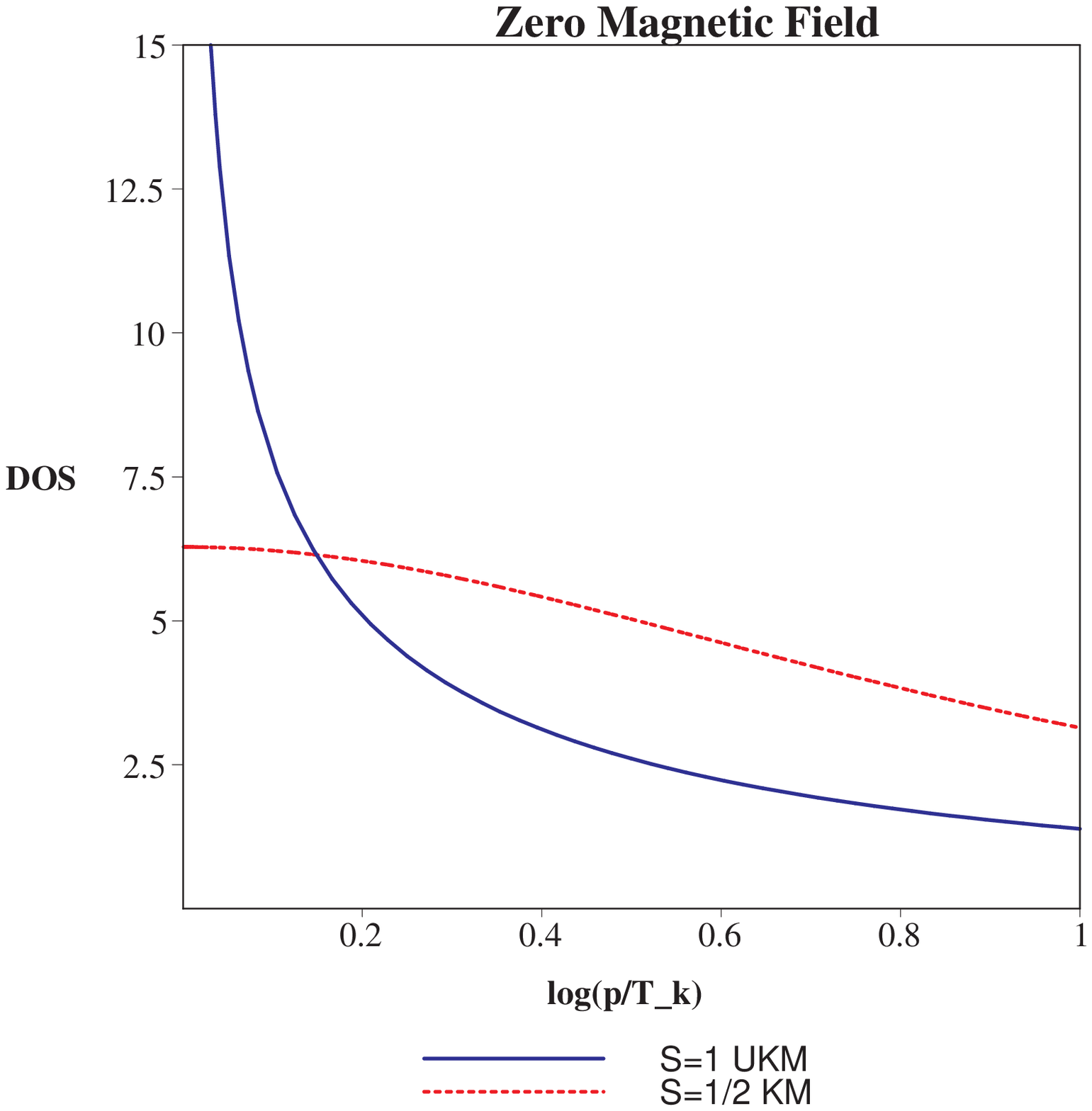}
\vspace*{0.25in}
\includegraphics[width=0.6\columnwidth]{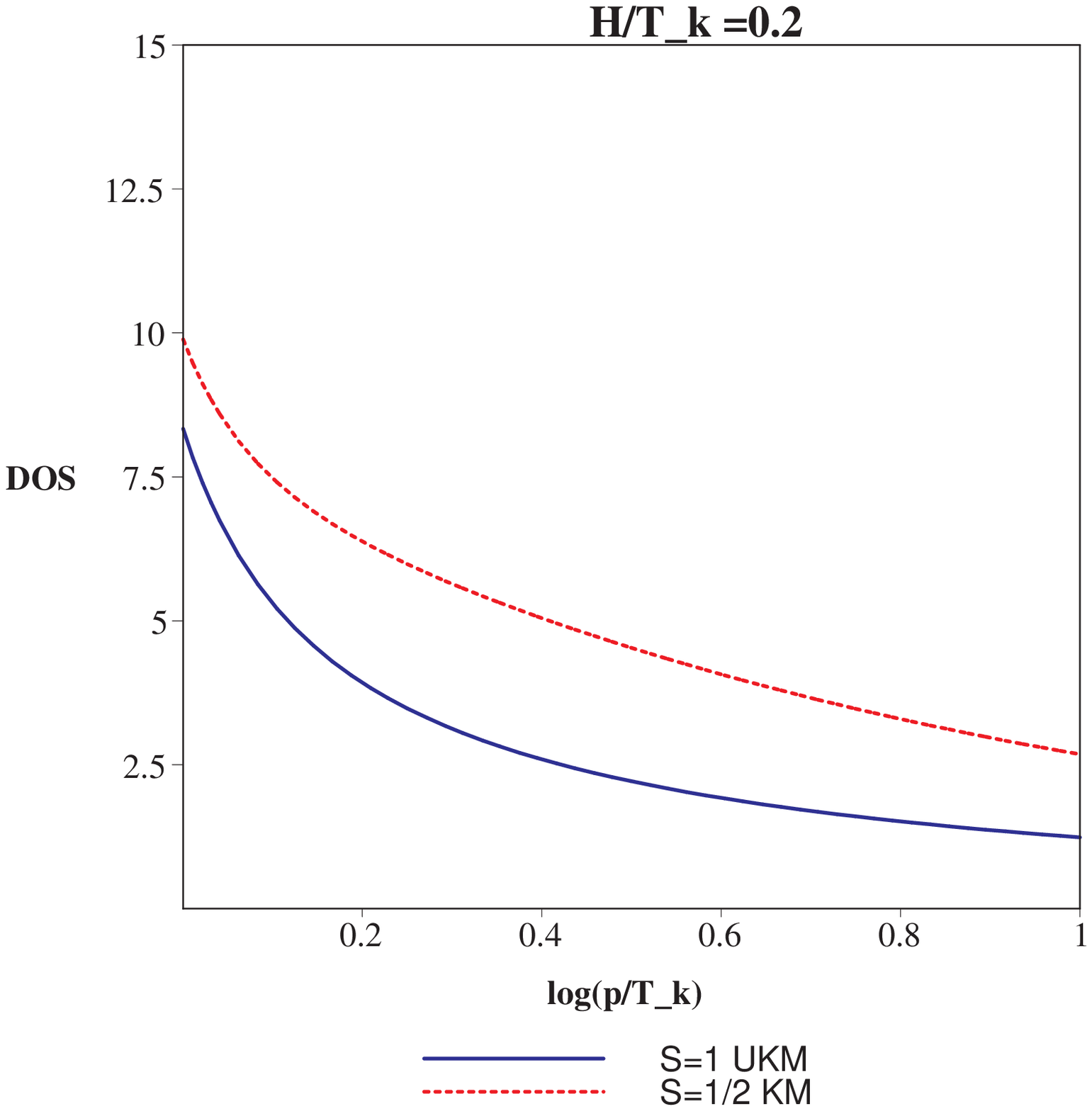}
\caption{\label{fig:BA_DOS} 
Fig.(a):  The impurity induced (spinon) DOS of the $S=1/2$ Kondo model and the $S=1$ underscreened Kondo models as a function of the 
logarithm of the quasi-particle  (spinon) energy. Figure (b): The same quantities in a finite magnetic field H.  
For $S=1/2$, the Impurity induced DOS is always finite. For the $S=1$ underscreened Kondo model, however, 
the DOS diverges in zero magnetic field as the energy of the excitation goes to zero. 
The presence of $H$ cuts off the  singularity of the DOS of the underscreened Kondo model.}
\end{figure}

From the Bethe-Ansatz solution, we can calculate the phase shift
$\delta_s(k,H)$ of a spinon with momentum $k$ when it is scattered off
the impurity in the presence of a magnetic field $H$.  The phase
shift, in turn, is intimately related to the DOS of spinons at the
impurity through the Friedel sum rule, which states that the spinon DOS,
$N_s(\omega)$, is proportional to the derivative of the phase shift
with respect to the energy ~\cite{langreth}. Note that as the energy
of the spinon is linear in its momentum we shall use the symbols
for momentum, $k$, and energy, $\omega$, interchageably (we
have chosen units where $v_F=1$, so $\omega=k$.)
\be
N_s(\omega,H)=\frac1\pi\frac{ d\delta_s (\omega,H)}{d\omega }
\ee

To calculate the phase shift, we place our physical system in a
finite ring of length $L$. The momentum $k$ of a free spinon 
will satisfy $ k= \frac{2\pi}{L}n$, but in the presence of a  impurity,
by definition,  the momentum will be shifted from its free value
by twice the phase shift
\be
k= \frac{2\pi}{L}n + \frac{2\delta_s(k=\omega,H)}{ L}\;.
\label{momentum}
\ee 
Since one can, using the Bethe-Ansatz solution, determine spinon
momenta to accuracy $O(1/L)$, the phase shift can be exactly
determined 
directly from the Bethe-Ansatz spectrum \cite{natan3, natan4}.

To solve for the spectrum of the UKM and
to determine the phase shifts, it is necessary to solve a set of
coupled integral-equations called the Bethe-Ansatz equations (BAE)~\cite{footnote2}.
The BAE are written in terms of the spin rapidities, $\Lambda$, and a
spinon 'magnetic field', $\Lambda_B$ (related to $H$, see later).  Each set of
$\{\Lambda\}$'s and $\Lambda_B$ which solve the BAE give rise to a set of
physical momenta, $\{k\}$, and physical magnetic field $H$.

In the thermodynamic limit, instead of examining specific solutions of
the BAE, it is sufficient to study the density of solutions.  Let
$\sigma(\Lambda)$ denote the density of solutions of the BAE in an
interval $d\Lambda$ 
(not to be confused with the scattering cross section). 
A spinon excitation corresponds to removing a
$\Lambda=\Lambda^h$ from the ground state, {\em i.e.}, to adding a
density of ``holes'', $\sigma^h(\Lambda)= \delta(\Lambda -
\Lambda^h)$ \cite{footnote3}. 
The ``hole'' position $\Lambda^h$, determines the spinon
momentum $k(\Lambda^h)$ and its phase shift 
$\delta_s(k(\Lambda^h),H)$.
It should be noted that the hole density is ``dressed'' by the back
flow of the Fermi sea, which corresponds to a small change in the
ground state density, $\Delta\sigma(\Lambda)$.  It is essential to
take this back flow into account when calculating the excitation
energy, $E= \sum_{j=1}^{N^e} \frac{2\pi}{L}n_j + D \int d\Lambda\; \sigma(\Lambda)
\lbc \Theta(2\Lambda-2) - \pi \rbc$.

In terms of these densities, the BAE can be written as
\begin{align}
\label{BAE}
\sigma(\Lambda) + \sigma_h(\Lambda) = f(\Lambda) - 
\int_{\Lambda_B}^{\infty} K(\Lambda - \Lambda^\prime)\sigma(\Lambda^\prime)d\Lambda^\prime \nonumber
\end{align}
 with
\begin{eqnarray}
f(x)&=&\frac{N^e}{\pi} \frac{c/2}{(c/2)^2+(x-1)^2}+ \frac{N^i}{\pi}
\frac{(cs)}{ (cS)^2 +x^2}, \nonumber\\
K(x) & = & \frac{1}{\pi}\frac{c}{c^2+x^2}\;, \nonumber \\
\Theta(x) &= &-2 \, {\tan}^{-1}({\frac{x}{c}}) \nonumber
\end{eqnarray}
\cite{under}.
Here $S$ is the spin of the impurity, $N^e$ is the number of electrons,
$N^i$ is the number of ``dilute'' impurities, and $c$ the coupling constant.
The coupling $c$ is related to the original coupling $J$, however, the precise relation 
between these two couplings 
depends on the specific scheme used to regularize the local interactions \cite{natan5}.
Using the chain rule,  we can write the spinon DOS as:
\be
N_s(\omega)=\frac{1}{\pi}\frac{ d\delta_s }{d\omega}= 
\frac{1}{\pi}\left( \frac{ d \omega}{d\Lam^h} \right)^{-1}
\frac{ d\delta_s}{d\Lam^h}
\label{eqDOS}
\ee
where $\omega$ is calculated from the expression for the energy. 
To proceed, we note that  the density of solutions in the presence of a 
spinon excitation,
$\sigma(\Lam, \Lam^h)$, can be written as
\be
\sigma(\Lambda, \Lambda^h) = \sigma_o(\Lambda)+ \Delta \sigma (\Lambda, \Lambda^h)
\ee
where $\sigma_o$ is the density in  the ground state (with no holes present) and $\Delta \sigma$ is the change
in the density due to the excitation (presence of the hole $\Lam^h$).  We can further divide $\sigma_0$ into two terms, 
$\sigma_{\rm el}$, the electron contribution to the groundstate and $\sigma_{im}$, the impurity
contribution to the ground state.  It is known that the derivative of the phase shift as a function 
$\Lambda^h$, $\frac{d\delta_s}{d \Lambda^h}$, is  precisely the impurity 
contribution of to ground-state density of solutions evaluated at $\Lam^h$, 
$\sigma_{im}(\Lambda^h)$  (see ~\cite{natan4}). Note that $\sigma_{im}(\Lambda)$ depends
only on the ground state and does not know about the presence of the spinon 
The information about the spinon in the DOS comes only through the spinon excitation energy, $\omega$. 
These observations greatly simplify the calculation of the DOS carried out in the appendix. 
Finally, it should be noted that since we are interested in the behavior around $H=0$, 
the results we present here are valid only for magnetic fields much 
smaller than the Kondo temperature
temperature, $H/T_k \ll 1$.

In the appendix we explicitly solve the BAEs and compute the
DOS $N_s(\omega)$. We find,
\begin{widetext}
\begin{align}
N_s(\omega)= \frac{1}{2\pi} \lba
\frac{1}{\omega + H^\prime} 
{\rm Re \,}[ \beta (S+ i\frac{1}{\pi}\log{((\omega+H^\prime)/T_k)})] 
+ \frac{H}{2\pi (\omega +H^\prime)^2} {\rm Re \,}[ \beta ( S+ i
\frac{1}{\pi}\log{(H^\prime/T_k)})] \rba
\label{DOS}
\end{align}
\end{widetext}
with $H^\prime =(\frac{e}{2\pi})^{1/2}H$ and ${\rm Re}\,[ \beta (x)] $ defined to be the real part of the function 
\be
\beta(x) = \frac{1}{2}\left( \psi(\frac{x+1}{2}) - \psi(\frac{x}{2}) \right)
\ee
and $\psi(x)$ the DiGamma function.

In Figure \ref{fig:BA_DOS}, the DOS versus energy is plotted for the
UKM. Notice that for the UKM, the DOS is singular
in the absence of a magnetic field. As a result, characteristics of 
quasiparticles are not analytic near the Fermi surface leading to singular
thermodynamical behavior. Note that the singular behavior is 
cut off by a finite field magnetic field.
To compare with numerical RG, we must explicitly calculate the phase
shift. To do so, we integrate the above expression with respect to
$\omega$ to get,
\begin{widetext}
\begin{align}
\delta(\omega, H) = \frac{\pi}{2} + \frac{1}{2i} 
\log{\left( \frac{ \Gamma(S+\frac{1}{2} + \frac{i}{\pi} 
\log{(\frac{\omega+H^\prime}{T_k})})
\Gamma(S- \frac{i}{\pi}\log{(\frac{\omega+H^\prime}{T_K})})}
{ \Gamma(S+\frac{1}{2} -\frac{i}{\pi} 
\log{(\frac{\omega+H^\prime}{T_k})})
\Gamma(S+ \frac{i}{\pi}\log{(\frac{\omega+H^\prime}{T_K})})}\right)}
-\frac{H^\prime}{\sqrt{2\pi e}(\omega+H^\prime)}{\rm Re \,}[ \beta ( S+ i
\frac{1}{\pi}\log{(H^\prime/T_k)})] \nonumber
\end{align}
\end{widetext}
The integration constant could be fixed by noting that the expression 
for the DOS is valid for any spin $S$ allowing us to compare it to a spin-1/2
calculation carried out in ref \cite{natan}.
As a further check, note that for $S=1/2$ and zero magnetic field,
the above expression can be simplified using various Gamma function
identities and yields 
\be
\delta_{S=1/2}(\omega)=\pi/2 -\tan^{-1}(\frac{\omega}{T_k})
\ee
in agreement with earlier calculations \cite{natan4}.

For small energies and magnetic field, the above expression can be simplified
using Stirlings approximation and yields
\bea
\frac{\delta_s(\omega, H=0)}{\pi} &=& 0.5 + ( S-\frac{1}{2}) 
\frac{1} {2\log{\frac{T_k}{\omega}}} + \ldots \nonumber \\
\frac{\delta_s(\omega=0, H)}{\pi} &=& 0.5 + ( S-\frac{1}{2}) 
\frac{1}{2\log{\frac{T_k}{H}}} + \ldots 
\label{phasesingular}
\eea

Thus, when $H=0$, the quasiparticle (spinon) phase shift approaches a
unitary value, a hallmark of FL.  However, as promised
earlier, it does it in a singular
manner.  Furthemore, note that
this singular behavior disappears for the ordinary Kondo model
when $S=1/2$.  For these reasons, we classify the UKM
as a Singular Fermi Liquid (SFL) state. This singularity has
interesting consequences for the phenomenological strong coupling
picture developed by Nozi\`eres for the $S=1/2$ Kondo model.

\section{Numerical Renormalization Group}

In this section we shall use Wilson's numerical renormalization group method to compute the magnetic 
field dependence of the  phase shift of the quasiparticles  and compare these numerical 
results with those of the  Bethe Ansatz  ~\cite{Wilson}. As we argued earlier,  although this is not true in 
general, at the Fermi energy the  phase shifts of the spinons  obtained from the Bethe Ansatz  should coincide with 
that of electrons. 

In Wilson's NRG technique one maps the original 
Hamiltonian of the impurity problem  to a semi-infinite chain 
with the magnetic impurity  at the end of the chain.
The hopping amplitude decreases  exponentially along the chain, 
$t_{n,n+1} \sim \Lambda^{-n/2}$, where
 $\Lambda\sim 3$ is a discretization parameter and $n$ labels the lattice 
sites along the chain. As a next step, one considers  the  
Hamiltonians $H_N$ of chains of length $N$, 
and diagonalizes them  iteratively to obtain the approximate 
ground state and the excitation spectrum of the infinite chain 
$$
\dots \to H_{N-1} \to H_N \to H_{N+1} \to \dots\;.
$$

The Hamiltonian $H_N$ in this series simply describes the spectrum of $H_L$, 
the original Hamiltonian,  
in a finite one-dimensional  box  of size $L\sim \Lambda^{N/2}$. The 
spectrum of $H_N$ is rather complicated in general, 
however, in the vicinity of a low-energy fixed point 
the finite size spectrum  $H_L$ becomes {\em universal}, implying that the spectrum of the fixed point Hamiltonian
\begin{equation}
H^* \equiv \Lambda^{N/2} H_N \sim {L\over 2\pi} \;H_L
\end{equation}
does not depend on the iteration number $N$ apart from an even-odd oscillation, due to the change of boundary 
conditions with $N$.

\begin{figure}
\includegraphics[width=0.8\columnwidth]{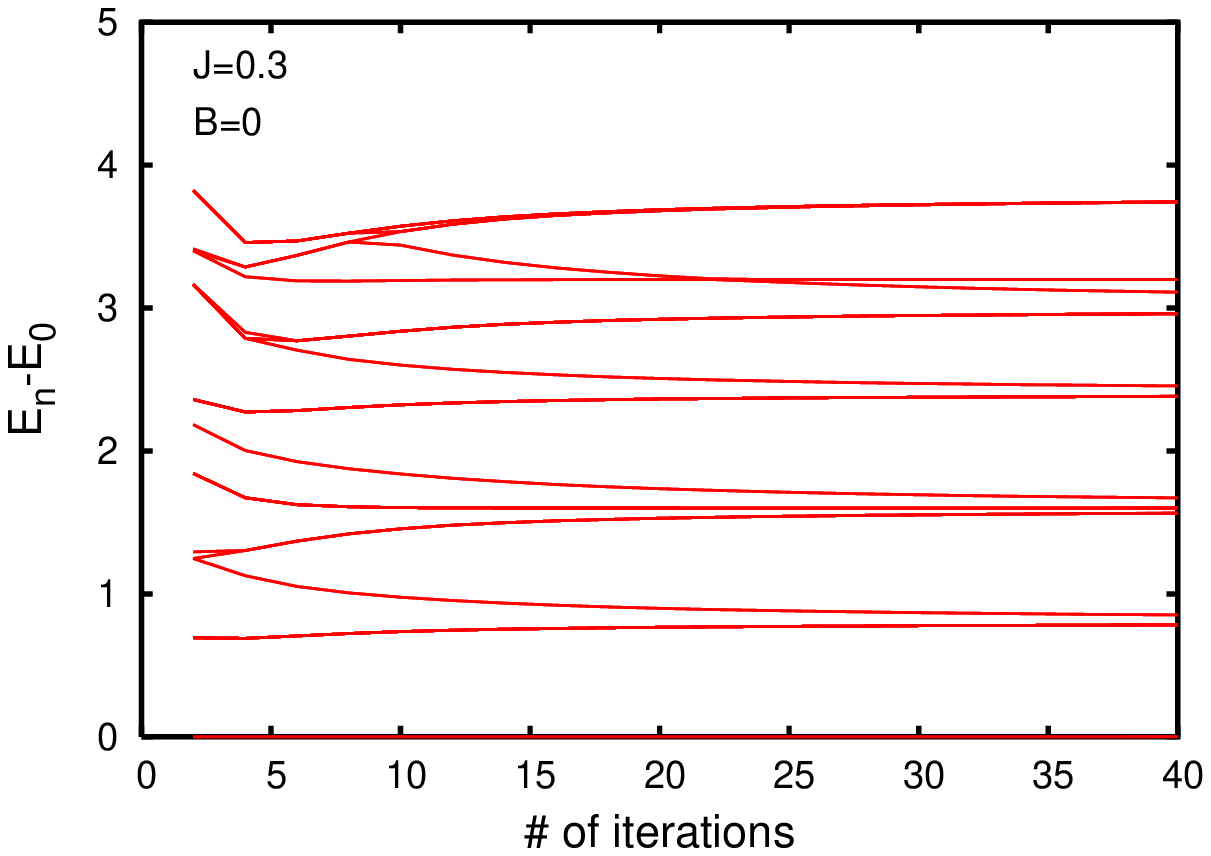}
\includegraphics[width=0.8\columnwidth]{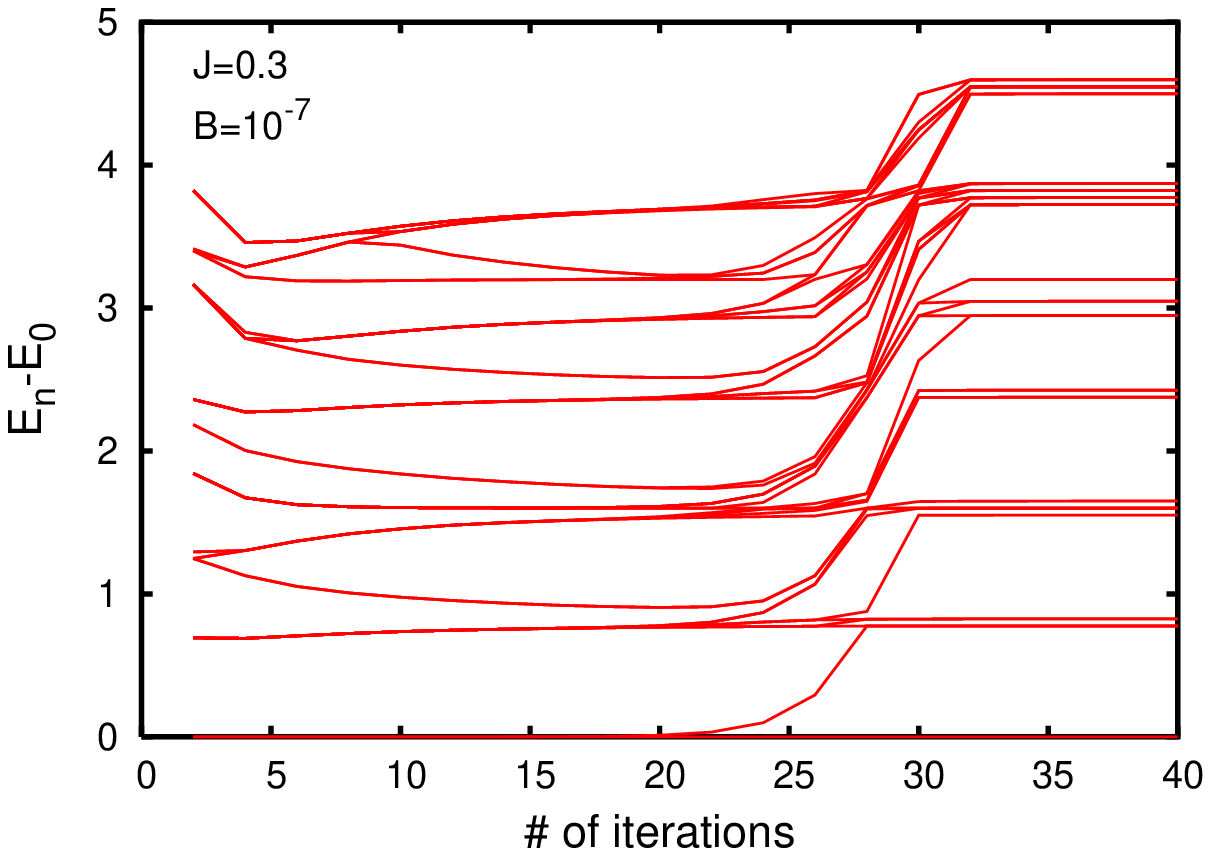}
\caption{
\label{fig:finitesize}
Finite size spectrum of the $S=1$ underscreened Kondo problem in the even sector in the absence (a) and presence 
(b) of a magnetic field. In the absence of a magnetic field the fixed point spectrum is that of a free Fermion field 
twisted by a phase shift $\pi/2$, and a residual spin $S^*=1/2$. 
 In a magnetic field a second scale appears below which the fluctuations of the 
residual spin $S^*=1/2$ are frozen, and the spectrum can be characterized by a single, field-dependent phase shift $\delta(H)$. 
}
\end{figure}

A typical finite size spectrum in zero magnetic field is shown 
in Fig.~\ref{fig:finitesize}. 
Only the spectra of  even iterations corresponding to 
periodic boundary conditions in the non-interacting problem are shown. 
For $N>5$ the excitation spectra approach very slowly ($\sim 1/N$) 
a universal spectrum. This universal spectrum is identical to that of a free residual spin $S^* = 1/2$ 
and the spectrum of the following Hamiltonian:
\begin{equation} 
H^* = {L\over 2\pi} 
\sum_{\sigma= \pm} \int_{-L/2}^{L/2}  dx \; \tilde \psi_\sigma^\dagger(x)  (-i \partial\tilde \psi_\sigma(x))\;, 
\label{eq:effective}
\end{equation}
where, in contrast to the original fields,  the free fermionic fields $\tilde \psi_\sigma(x)$ obey now {\em antiperiodic} boundary conditions:
\begin{equation}
\tilde \psi_\sigma(-L/2) =  -\tilde \psi_\sigma(L/2) \;.
\end{equation}

Thus in the absence of a magnetic field fermions at the Fermi energy simply acquire a phase shift $\pi/2$. As a consequence, 
the spectrum of Eq.~(\ref{eq:effective}) is gapped for a finite system size,  
and  the ground state of the system is only two-fold degenerate 
due to the presence of the residual spin $S^*$.  As shown in Fig.~\ref{fig:finitesize}.b, in the presence of a small 
magnetic field $H$ a new scale $\sim H$  emerges, below which the fluctuations of the residual spin are frozen out, 
and the ground state degeneracy is lifted. Below this scale the spectrum 
can be described simply by Eq.~(\ref{eq:effective}) with the modified boundary conditions  
\begin{equation}
\tilde \psi_\sigma(-L/2) =  -e^{-i 2\delta_\sigma(H)} \tilde \psi_\sigma(L/2) \;,
\end{equation}
where $\delta_\sigma(H)$ denote field-dependent phase shifts. 
Note that these phase shifts are 
the phase shifts of {\em charged excitations}, {\em i.e.}, from the NRG spectrum we determine directly the 
phase shifts of the electrons at the Fermi energy.

\begin{figure}
\includegraphics[width=0.9\columnwidth]{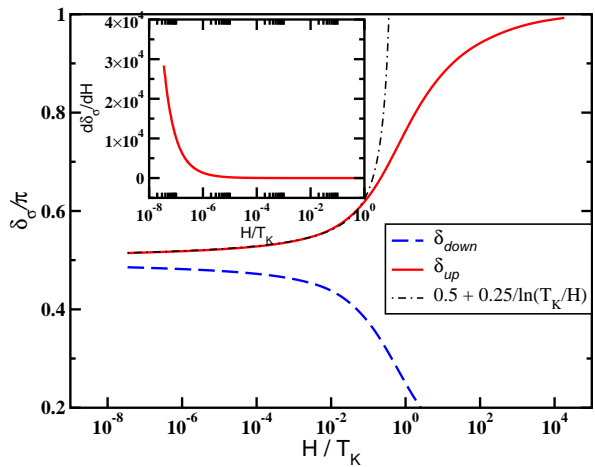}
\caption{
\label{fig:phaseshift_NRG}
Magnetic field dependence of the phase shifts extracted from the NRG finite size spectrum. 
The phase shifts scale to $\pi/2$ as $\sim 1/{\rm ln}(T_K/|H|)$.
Inset: The derivative of the phase shift diverges as $1/|H|$ for $|H|\ll T_K$.
}
\end{figure}

We can thus determine the magnetic field dependence of the phase 
shifts directly  
from the NRG spectrum. As shown in shown in Fig.~\ref{fig:phaseshift_NRG}, 
the phase shifts approach $\pi/2$ as $0.29/{\rm ln}(T_K/H)$ 
in good agreement with the Bethe-Ansatz result for $S=1$ Eq.(\ref{phasesingular}). 
In the inset of Fig.~\ref{fig:phaseshift_NRG} we plotted the 
derivative of the phase shift too, that we computed by numerically differentiating the  NRG results. 
This derivative is proportional to the  quasiparticle density of states at the Fermi level, and  indeed 
diverges approximately as $\sim 1/H$ for $H\to 0$.

\section{The Breakdown of  Nozi\`eres' Fermi Liquid Picture for the UKM}

In his seminal papers \cite{noz},  Nozi\`eres argued that one could perform
a ``Fermi Liquid expansion of phase shifts'' at strong coupling.
He argued that since the impurity is frozen into a singlet at strong coupling, 
the only remaining degrees of freedom in the problem were those of the Fermi liquid.
He showed that all the physics could be captured by examining the phase
shifts of quasi-particles as they pass the impurity. We shall now argue that 
this picture is valid for RFL but fails in the case of SFL.

Nozi\`eres' prescription to describe a Fermi liquid is to {\em assume} that the phase shift for 
a quasiparticle of energy $\omega$ and spin  $\sigma$  has the general form 
\be
\delta_\sigma(\omega) = \delta_\sigma[\omega, \{n_{\sigma'}(\omega') \}]\;, 
\ee
where $\{n_{\sigma'}(\omega') \}$ denotes the occupation number of all other quasiparticle states. 
It is not clear from Nozi\`eres original paper how exactly the phase shift can be defined for a particle of finite energy, 
which  scatters generically inelastically off the impurity. Implicitly, Nozi\`eres' prescription assumes, 
that sufficiently close to the Fermi surface the inelastic scattering of a quasiparticle of energy $\omega$ is suppressed 
as $\sim \omega^2$, and thus  quasiparticles are indeed well-defined. With this assumption, and assuming further that 
in the  strong coupling fixed point everything is {\it analytic} near the Fermi surface
 one  can  proceed and  expand the phase shift in powers of $\omega$ and the change of quasiparticle occupation number,
$\delta n$, as
\begin{eqnarray}
\delta_\sigma(\omega) &=& \delta_0(\omega) 
+ \sum_{\omega^\prime, \sigma^\prime} \phi_{\sigma, \sigma^\prime}
(\omega , \omega^\prime) \delta n_{\sigma^\prime}(\omega^\prime) \;,
\nonumber \\
\delta_0(\omega) &=& \delta_0 + \alpha \omega + \beta \omega^2\;, 
\end{eqnarray}
where for the sake of simplicity we assumed $H=0$. These equations are the main constituents 
of Nozi\`eres' Fermi liquid theory. The assumption that $\delta_0(\omega)$ is analytical in 
$\omega$ implies that the impurity-induced DOS remains finite at the Fermi energy with $\alpha \sim 1/T_K$.

Our Bethe Ansatz solution, however, shows that in the absence of a
magnetic field the spinon phase shifts take the form 
\bea
\delta_s(\omega) = \frac{\pi}{2} + \gamma \frac{\sign(\omega)}{ \ln \left(
\frac{T_K}{\omega}\right)} + \dots\;, 
\eea 
leading to the singular
density of states for the spinon excitations shown in
Fig. \ref{fig:BA_DOS},
\bea N_s(\omega) = \frac{1}{\pi} \frac{\partial
\delta_s}{\partial \omega} = \frac{\gamma}{\pi \vert\omega\vert \left(\log \left(
\frac{T_K}{\omega}\right)\right)^2}\;.  
\eea 
As a results the conventional
Fermi liquid expansion of the phase shift can not be carried out.

Another essential feature of the Nozi\`eres Fermi liquid approach, is
the assumption of adiabaticity - that the excitations of the
interacting system can be mapped onto the excitations of a
corresponding non-interacting impurity problem. Since the interacting
and non-interacting systems contain the same quasi-particles, the
difference between the two situations can only be due to scattering by
a one-particle potential.

We are thus lead to ask whether there is adiabaticity in the UKM. In
light of the above observation, we can phrase the question in an
alternative manner - is there any non-interacting scattering potential
that can give rise to the observed energy-dependent spinon phase
shift?  In a conventional impurity scattering problem, the scattering
potential and the phase shift are related by the relation \bea
\delta(\omega) = \tan^{-1}[- \pi V(\omega ) \rho] \eea where
$V(\omega) $ is the bare scattering potential at energy $\omega$
~\cite{Nozphase}, so that \bea V(\omega ) = - \frac{1}{\pi\rho }\tan
\delta(\omega) \eea In the Nozi\`eres expansion, we have \bea \delta =
\frac{\pi}{2} + \alpha \omega \eea so that the corresponding potential
is given by \bea V(\omega) = \frac{1}{\pi\rho \alpha
}\frac{1}{\omega}.
\label{eq:V(omega)}
\eea
A  $1/\omega$ phase shift indicates the formation of single
bound-state inside the Kondo resonance. 
In fact, the scattering potential (\ref{eq:V(omega)}) is the same as that 
of a simple resonant level model with  a resonance of width
$\Gamma\sim \alpha^{-1} \sim T_K$ positioned right at the Fermi energy, 
$\epsilon_d = 0$, implying that 
we can indeed map the excitations of the
fluid onto a non-interacting Anderson impurity model.

If we now carry out the same procedure on the phase shift of the
UKM, we find that 
\bea
V^*(\omega) = \frac{1}{\pi\rho \gamma }\left[
\ln \left(\frac{T_K}{\omega}\right)
\right]\sign(\omega)
\eea


This singular elastic scattering potential
can not be replaced by a simple scattering pole, but would require an 
singular distribution of 
non-interacting scattering resonances for its correct description. 
In this way, we see that the singular Fermi liquid of the underscreened
model can not be obtained from the adiabatic evolution of a simple,
non-interacting impurity model.

\section{Conclusion}

We end our discussion by remarking on some interesting lines for future
research. The UKM model discussed here is isotropic, and the characteristic
scale for the field-tuned Fermi liquid is, up to a logarithm, a linear function of the magnetic field.  
It may be particularly interesting in future work to examine the properties of the anisotropic UKM
\begin{eqnarray}
&&H_{UKM} = -i \int dx \, \psi^\dag(x) \partial \psi(x)\\ 
 &+ &J_z\lba \psi^{\dag}(0) {\sigma}_z \psi(0)\rba {S}_z +
J_{\perp} \lba \psi^{\dag}(0) {\sigma}_{\perp} \psi(0)\rba {S}_{\perp}
\;. \nonumber 
\end{eqnarray}
The low-temperature physics of this model maps onto an anisotropic
ferromagnetic Kondo model at strong coupling, where the physics
is described by a line of fixed points \cite{TG}. In this problem, we expect
that the linear specific heat coefficient will diverge with an exponent
that depends on the degree of anisotropy, 
\bea
\frac{C_v(T)}{T} \sim \frac{1}{T^{\alpha(J_z/J)}}\phi(\frac{T}{B})
\eea
where $\phi(x)$ is a scaling function. 
This kind of behavior has recently been seen\cite{gegenwart, Custers} in the field-tuned 
QCP of YbRh$_2$Si$_2$, and the anisotropic 
underscreened Kondo model may provide an interesting point of comparison with the
field tuned physics in anisotropic quantum critical systems.

Finally, in the spirit of the Nozi\`eres picture, Affleck and Ludwig
have analyzed the low energy behavior of Kondo impurity models in the
framework of boundary conformal field theory (BCFT) ~\cite{Affleck}.
In this method, the various fixed points correspond to different
conformally invariant boundary conditions. Although the over-screened
and exactly-screened Kondo models were analyzed in great detail, the
UKM were never properly examined, and it is still an open question
how to incorporate  the SFL behavior of the UKM we have found in terms of
 BCFT.  

\section{Acknowledgements}
We have benefited from discussions and email exchanges with many people, including 
E. Boulat and I. Paul. 
This work has been supported by  NSF through grants DMR 9983156 and 
DMR 0312495, 
by the Bolyai foundation, the NSF-MTA-OTKA Grant No. INT-0130446,  
and Hungarian grants No. OTKA T038162, T046303, and T046267
and the EU  `Spintronics' RTN HPRN-CT-2002-00302.

\appendix
\section{Appendix 1: Calculation of DOS}

In this appendix, we outline the calculation of the DOS for a spin-S, single
channel Kondo model. As discussed in the text (\ref{eqDOS}), the DOS is given by
the derivative of the phase shift with respect to the energy excitation
which can be rewritten as
\be
\frac{ d\delta_s }{d\omega } = \left( \frac{ d \omega_s}{d\Lam^h} \right)^{-1} 
\frac{ d\delta_s}{d\Lam^h} =  \left( \frac{ d \omega_s}{d\Lam^h} \right)^{-1} \s_{im,B}(\Lam^h)
\ee
where $\omega$ is the excitation energy of the spinon and $\Lam^h$ is the 
hole induced in the spin-rapidity due to the presence of a spinon.
Hence, the calculation naturally divides into two parts, calculation
of $\sigma_{im}(\Lam)$ and the calculation of the excitation energy $\omega$.

We first concentrate on calculating $\sigma_{im}(\Lam)$. Since, $\sigma_{im}(\Lam)$ is not 
affected by the presence of the spinon, we can calculate  $\sigma_{im}(\Lam)$ directly
in the ground state. The starting point for the calculation is the equation for the ground state 
in the presence of a magnetic field ~\cite{footnote2}
\begin{equation}
\label{gseq}
\s_B (\Lam) + \int_{\Lam_B}^\infty d\Lam^\prime
  K(\Lam - \Lam^\prime) \s_B (\Lam^\prime)= f(\Lam)
\end{equation}
As explained in the text, $\Lam$ is the spin-rapidity and $\Lam_B$ is a parameter related to the 
physical magnetic field, $H$, by the relation 
$\frac{H}{T_k}= \left( \frac{2}{\pi e} \right)^{1/2}e^{\pi/c \Lam_B}$ where $T_k$ the Kondo temperature.

Shifting  the limits of the integral and defining
$\rho(\Lam) = \s_B(\Lam +\Lam_B)$ the above can be rewritten as
\be 
\label{Bground}
\rho(\Lam) + \int_0^\infty d\Lam^\prime 
K(\Lam - \Lam^\prime)\rho(\Lam^\prime) = f(\Lam + \Lam_B) 
\ee

This equation can be solved using the Weiner-Hopf technique.
This technique relies on separating all expressions into
a sum of expressions that have singularities only in either
the upper of lower half-plane. After separating all the expressions,
one can equate those expressions that are analytic in each half-plane
separately.Most of the manipulations in the appendix
are performed in order to facilitate this separation.
To proceed, define $\rho_{\pm}(\Lam)= \theta(\pm\Lam)\rho(\Lam)$. Then, 
($\ref{Bground}$) becomes
\be
\rho_+(\Lam) + \rho_-(\Lam) + \int_{-\infty}^{\infty}d\Lam^\prime 
K(\Lam-\Lam^\prime) \rho_+(\Lam^\prime) = f(\Lam +\Lam_B)
\ee

Taking the Fourier Transform of the above and noting that
the integral is a convolution one has the equation
\be
\label{rhoground}
\tilde{\rho}_+(p)\lba 1+ \tilde{K}(p) \rba + \tilde{\rho}_-(p) = 
\tilde{f}(p) e^{ip\Lam_B}
\ee
One now wants to separated the terms in the above equation 
into functions with singularities only in the upper/lower
half plane. Hence, one rewrites the above equation as 
\be
1+\tilde{K}(p)=\frac{K_+(\frac{cp}{2\pi})}
{K_-(\frac{cp}{2\pi})}
\ee
where $K_+(K_-)$ have singularities only in the upper(lower) half plane.
Explicitly, one has
\bea
&K_+(q)= K_-^{-1}(-q)& \nonumber \\
&=
\frac{(2\pi)^{1/2}}{\Gamma(\frac{1}{2}+iq)}\exp\left[-iq \left[1+\frac{i\pi}{2}-\log{(-q+i\epsilon)}\right]\right]
\eea
Equation ($\ref{rhoground}$) can be rewritten as
\begin{eqnarray}
\label{posnegground}
\tilde{\rho}_+(p) K_+(\frac{cp}{2\pi})&+ \tilde{\rho}_-(p)
K_-(\frac{cp}{2\pi}) \nonumber \\
&=  K_-(\frac{cp}{2\pi}) \tilde{f}(p)e^{ip\Lam_B}& \Lambda_B\geq 0 \nonumber \\
&= K_+(\frac{cp}{2\pi}) \tilde{g}(p)e^{ip\Lam_B}& \Lambda_B\leq 0
\end{eqnarray}
where 
\bea
\tilde{f}(p) &=& N^e e^{-c/2|p|} + N^i e^{-cs|p|}  \nonumber \\
\tilde{g}(p) &=&\frac{\tilde{f}(p)}{1+\tilde{K}(p)} =  
\frac{N^e e^{ip} + N^i e^{(S-1/2) |p|}}{2 {\cosh \lba\frac{c}{2}p \rba}}
\label{defg}
\eea
 
Inverse Fourier transforming the above expression for $\tilde{g}(p)$, one has
\be
g(\Lam) = \frac{1}{2c}
\frac{N^e}{{\cosh \lba \frac{\pi}{c}(\Lam -1)\rba}} +\sum_{k=0}^{\infty}
N^i \frac{(-1)^k c(S+k)}{c(S+k)^2 + \Lam^2}.
\ee
Since we are interested in small magnetic fields,
we concentrate on the $\Lambda_B\leq 0$ solutions.
In order to separate the right hand side of (~\ref{posnegground}) into
parts that have singularities only in one half-plane, it is useful to rewrite
$\tilde{g}(p)$ in an alternative way.
To do this, we follow ~\cite{natan} and Laplace transform the above expression for $g(\Lam)$ 
to get
\begin{widetext}
\be
g(\Lam) =\sum_{k=0}^{\infty}(-1)^k \lba \frac{N^e}{c}
e^{-\frac{\pi}{c}(2k+1)|\Lam -1|}+ N^i\int_0^\infty dt {\sin \lba c(S+k)t \rba}
e^{-|\Lam|t} \rba
\ee
\end{widetext}

Since we are interested in $\tilde{g}(p)$, we  Fourier
transform each amplitude of the Laplace transform separately.
Define $\tilde{g}_t(\Lam)= e^{-|\Lam|t}$. Its Fourier transform is given by 
\be
\tilde{g}_t(p) = i\lba \frac{1}{p+it} - \frac{1}{p-it} \rba
\ee
Notice that each term in (~\ref{posnegground}) can be
written as an sum or  integral of terms of the form
$K_+ \lba \frac{cp}{2\pi}\rba \tilde{g}_t(p) e^{ip\Lam_B}$. We want to write each 
term as a sum of functions that have singularities 
solely in the  upper/lower half-plane. Define $\tilde{\rho}_\pm (p,t,\Lam_B)$
to be 
\be
\tilde{\rho}_\pm (p,t,\Lam_B)= \frac{q_\pm^{''}(p, t, \Lam_B)} { K_\pm(p) }
\ee
where
\begin{widetext}
\begin{align}
q_+^{''}(p, t, b) =&(-i) K_+ \lba \frac{cp}{2 \pi} \rba \frac{e^{ip\Lam_B}}
{p-it} + i \lbc K_+ \lba \frac{cp}{2\pi} \rba e^{ip\Lam_B} - K_+ \lba
\frac{-ict}{2 \pi} \rba e^{\Lam_Bt} \rbc \frac{1}{p+it} \\
q_-^{''}(p, t, b) =& iK_+\lba \frac{-ict}{2\pi} \rba \frac{e^{\Lam_Bt}}{p+ it}
\end{align}
Note that $\tilde{\rho}_+ (p,t,\Lam_B)$ [$\tilde{\rho}_- (p,t,\Lam_B)$] has
singularities only in the upper (lower) half p-plane.
Then,
\be
\rho_\pm(p) = \sum_{k=0}^\infty (-1)^k \lbc \frac{N^e}{c} 
\tilde{\rho}_\pm \lba p, t=(2k+1)\frac{\pi}{c}, \Lam_B-1 \rba + \frac{N^i}{\pi}
\int_o^\infty dt \, {\sin \lba c(S+k)t \rba}\tilde{\rho}_\pm \lba p, 
t,\Lambda_B\rba \rbc
\ee
\end{widetext}
Since we want to calculate $\sigma_{im}$, the  density of states at the impurity,
we can simply concentrate only on the part of the above expression that is 
proportional to $N_i$. 
Inverse transforming the above expressions and adding them one has
\begin{widetext}
\bea
\rho_{im}(\Lam)&=& N^i \sum_{k=0}^\infty (-1)^k 
\int_0^\infty dt \,{\sin \lba c(s+k)t \rba} \frac{1}{2 \pi i} 
\lbc \int_{-\infty}^{\infty} dp \,\frac{ e^{ip(\Lam +\Lam_B)}}{p-it}
-\frac{e^{ip(\Lam +\Lam_B)}}{p+it} \rbc \nonumber \nonumber \\
&+& N^i\int_0^\infty dt \, {\sin \lba ct(S+k) \rba} \frac{1}{2 \pi i} 
\lbc \int_{-\infty}^{\infty} dp \, e^{-ip\Lam}e^{\Lam_Bt}K_+(ict/2\pi)
\frac{K_+(cp/2\pi)-K_-(cp/2\pi)}{p- it} \rbc
\eea
\end{widetext}
From now on we will be concerned exclusively with the impurity
portion
Noting that $\s_{B,im}(\Lam+\Lam_B) = \rho_{im}(\Lam) $ one can write
\be
\s_{im,B}(\Lam) = \s_{im,H=0}(\Lam) + \s_{im, H\neq 0}(\Lam) \nonumber 
\ee
where we have separated the impurity density into a zero-field, $\s_{im,H=0}(\Lam)$
and finite-field contribution,  $\s_{im, H\neq 0}(\Lam)$. Explicitly, these are given
by
\begin{widetext}
\begin{eqnarray}
\s_{im, H=0}(\Lam) &=& N^i \sum_{k=0}^\infty (-1)^k 
\int_0^\infty dt \,{\sin \lba c(S+k)t \rba} \frac{1}{2 \pi i} 
\lbc \int_{-\infty}^{\infty} dp \,\frac{ e^{ip\Lam}}{p-it}
-\frac{e^{ip\Lam}}{p+it} \rbc \nonumber \\
&=&  \sum_{k=0}^\infty \frac{(-1)^k c(S+k)}{\lbc c(S+k)\rbc^2 + \Lam^2}
\end{eqnarray}
and 
\bea
\label{NoHsigma}
\s_{im, H \neq 0}(\Lam) = N^i \sum_{k=0}^\infty (-1)^k 
\int_0^\infty dt \, {\sin \lba ct(S+k) \rba} \frac{1}{2 \pi i}
\lbc \int_{-\infty}^{\infty} dp \, e^{-ip(\Lam-\Lam_B)} e^{\Lam_Bt}K_+(ict/2\pi)
\frac{K_+(cp/2\pi)-K_-(cp/2\pi)}{p- it} \rbc 
\eea
\end{widetext}
For what follows, we assume $\Lambda_B\ll 0$ or the physical magnetic field
$H /T_k\ll 1$. In equation ($\ref{NoHsigma}$), the integral over t is dominated
by t very close to zero. Hence, we can substitute t=0 in $K_+$ integrate
over t to get the expression
\begin{widetext}
\begin{multline}
\s_{im, H \neq 0}(\Lam) = N^i \sum_{k=0}^\infty \frac{ (-1)^k c(S+k)}
{c^2(S+k)^2 +\Lam_B^2}\times\frac{1}{2 \pi i}
\lbc \int_{-\infty}^{\infty} dp \, e^{-ip\frac{\pi}{c}(\Lam-\Lam_B)}
\frac{K_+(p/2)-K_-(p/2)}{p- i\epsilon} \rbc 
\end{multline}
\end{widetext}
We close the contour below and noting that the poles in $K_{-}$ arise
from Gamma function one has in the universality limit:
\begin{equation}
 \s_{H \neq 0}(\Lam) = N^i \sum_{k=0}^\infty \frac{ (-1)^k c(S+k)}
{c^2(S+k)^2 + \Lam_B^2}\times\frac{e^{-\frac{\pi}{c}(\Lam - \Lam_B)}}{\sqrt{2\pi e}}
\end{equation}
Hence, for $H<<T_k$, the expression for the impurity density $\s_{im, B}(\Lam)$
in the presence of a magnetic field is given by
\begin{widetext}
\begin{equation}
\s_{im, B}(\Lam)  =   N^i \sum_{k=0}^\infty \lbc\frac{ (-1)^k c(S+k)}
{c^2(S+k)^2 + \Lam^2}+ \frac{ (-1)^k c(S+k)}
{c^2(S+k)^2 + \Lam_B^2}\times\frac{e^{-\frac{\pi}{c}(\Lam - \Lam_B)}}{\sqrt{2\pi e}}\rbc
\end{equation}
\end{widetext}

Having calculated, $\s_{im, B}(\Lam)$, in order to calculate the DOS, we must now calculate  
$\left( \frac{ d \omega}{d\Lam^h} \right)$ where $\omega$ is the excitation energy of the 
spinon excitation and depends explicitly on the hole position $\Lam^h$.
To calculate spinon energy one starts with the BAE equation in the presence of a hole
\be
\label{exeq}
\s_B(\Lam) + \delta(\Lam-\Lam^h) + \int_{\Lam_B}^\infty K(\Lam - \Lam^\prime)
\s_B(\Lam^\prime) d\Lam^\prime = f(\Lam)
\ee
Define $\Delta \s_{B,\Lam^h}(\Lam)$, the change in the spin-rapidity due
to the hole,  by
\be
\s_B(\Lam) + \d (\Lam -\Lam^h) =\s_{B,gs}(\Lam) +
\Delta \s_{B,\Lam^h}(\Lam) .
\ee

Substituting the above definition into ($\ref{exeq}$) and using ($\ref{gseq}$) one has,
\be
\Delta \s_B(\Lam) + \int_{\Lambda_B}^{\infty} \Delta\s_B(\Lam^\prime) K(\Lam -\Lam^\prime) d\Lam^\prime = 
K(\Lam - \Lam^\prime) \nonumber
\ee
Shifting the integral, taking the Fourier Transform, and rewriting in terms of $K_\pm(p)$ 
one has
\begin{widetext}
\be
 e^{ip\Lam_B}\tilde{h}(p)=K_+(\frac{cp}{2\pi})\Delta \tilde{\rho}_+(p) +  K_-(\frac{cp}{2\pi})\Delta \tilde{\rho}_-(p)
= \frac{e^{-c|p|}e^{-ip(\Lam^h-\Lam_B)}}{1+e^{-c|p|}}= e^{-ip(\Lam^h-\Lam_B)}\sum_{k=o}^{\infty}(-1)^k e^{-c(k+1)|p|}
\ee 
\end{widetext}
where we have defined a new function $h(\Lambda)$ whose Fourier transform
is given by the expression $\tilde{h}(p)$ above.
Explicitly, 
\be
h(\Lam) = \frac{1}{4} \sum_{k=1}^{\infty} (-1)^{k+1}
\frac{ck}{(ck)^2+(\Lam -(\Lam^h) ^2)}
\ee

In analogy with the manipulations we used for the second term in
(\ref{defg}) when calculating the DOS, we write $h(\Lam)$ as a Laplace 
transform, Fourier transform each Laplace 
component, and then equate functions with  singularities only in
the upper and lower half planes to get  
\begin{widetext}
\begin{equation}
\label{sigmaminus}
\Delta \tilde{\s}_-(p) = \sum_{k=0}^{\infty}(-1)^k \int_0^\infty dt \;{\sin (c(k+1)t)}
 i\frac{K_+ \lba \frac{-ict}{2\pi} \rba e^{-ip\Lam_B}}
{K_- \lba \frac{cp}{2\pi} \rba} \frac{e^{-(\Lam^h-\Lam_B)t}e^{\Lam_Bt}}{p+it}
\end{equation}
and
\be
\label{sigma+p}
\Delta  \tilde{\sigma}_+(p) = \frac{1}{4}\sum_{k=1}^{\infty}(-1)^{k+1}
\int_0^{\infty} dt\;{\sin (ckt)} (-i) 
\lbc \frac{e^{-ip\Lam^h}}{p -it} - \frac{e^{-ip\Lam^h}}{p+it} 
+\frac{e^{-(\Lam^h-\Lambda_B)t}e^{-ip\Lam_B}}{p+it}\frac{K_+(-ict/2\pi)}{K_+(cp/2\pi)}
\rbc
\ee

The excitation energy is given by ($D=N^e/L$ is the bandwidth)
\bea
 \omega(\Lam^h) = D\int_{\Lam_B}^{\infty}d\Lam \, \Delta \s_+(\Lam) \lbc \theta(2\Lam-2) - \pi \rbc 
-H \int_{-\infty}^{\Lambda_B} d\Lam \Delta\s_-(\Lam)
\end{eqnarray}
To proceed, note that the second term can be rewritten as
\be
H \int_{-\infty}^{\Lambda_B} d\Lam \Delta\s_-(\Lam) 
= \int_{-\infty}^0 d\Lam \Delta \rho_-(\Lam) 
= \int_{-\infty}^{\infty} d\Lam \Delta \rho_-(\Lam) 
= H \Delta \tilde{\rho}_-(0) 
\ee
For, very small $H \Delta \tilde{\rho}_-(0)$ tends to $H^\prime
=H(\frac{e}{2\pi})^{1/2}$. 
After some  manipulations, the first term can be written

\bea
\omega(\Lam^h) &=&\int_{0}^{\infty}dp - \frac{{\sin(p\Lam^h)}}{(1+e^{cp})p}e^{-cp/2}+ 
\frac{\pi}{2} \nonumber \\
&+& \frac{1}{4}\sum_{k=1}^{\infty}(-1)^{k+1}
\int_0^{\infty} dt{\sin (ckt)} \frac{e^{-(\Lam^h-\Lambda_B)t}e^{-ip\Lambda_B}e^{-c|p|/2}}
{p(p+it)}\frac{K_+(-ict/2\pi)}{K_+(cp/2\pi)}-H^{\prime}
\eea
\end{widetext}
The sum of the first two terms in  the expression above is the
the expression for the excitation energy for zero field.
Hence, we know that in the universality limit
\begin{widetext}
\begin{multline}
\omega(\Lam^h)= 2T_0 e^{\pi/c\Lam^h}+
\sum_{k=1}^{\infty}(-1)^k \int_0^\infty dt {\sin (ckt)}
K_+ (-ict/2\pi)  e^{-\Lam^h t}e^{\Lambda_Bt}
\lbc \frac{H}{\sqrt{2}t} + \int_{-\infty}^{\infty} dp
 \frac{e^{-ip\Lambda_B}e^{-c|p|/2}}{p(p+it)K_+(cp/2\pi)} \rbc -H^{\prime}
\label{exenergy}
\end{multline}
\end{widetext}
For $B\ll 0$ (very small fields), we can ignore the second term giving us the 
result:
\be
\frac{d\omega}{d\Lam^h}= T_0\frac{\pi}{c}e^{\pi/c\Lam^h}
\ee
Hence we see that the DOS of spinons at the impurity is given by:
\begin{widetext}
\begin{equation}
 N_s(\omega(\Lam^h))=  \frac{c}{\pi^2 T_K} \sum_{k=0}^\infty 
\lbc e^{-\pi/c \Lam^h}\frac{ (-1)^k c(S+k)}{c^2(S+k)^2 + {\Lam^h}^2}
+\times\frac{e^{-\pi/c(2\Lam^h-\Lambda_B)}}{\sqrt{2\pi e}}\frac{ (-1)^k c(S+k)}
{c^2(S+k)^2 + \Lambda_B^2}\rbc
+\ord(H)
\end{equation}
where it is implicitly assumed that $\Lambda^h > \Lambda_B$
We can rewrite the above equation
\begin{equation}
N_s(\omega(\Lam^h))= \frac{1}{2\pi^2 T_k} \lba
e^{-\frac{\pi}{c}\Lambda^h} {\rm Re \,}[\beta(S+ i\Lambda^h/c)] 
+ \frac{e^{-\pi/c(2\Lam^h-\Lambda_B)}}{\sqrt{2\pi e}} {\rm Re \,}
[\beta(S+ i\Lambda_B/c)] \rba
\end{equation}
where Re [f] denotes the real part of f and 
\be
\beta(x) =\frac{1}{2}\left( \psi(\frac{x+1}{2}) - \psi(\frac{x}{2}) \right)
\ee
Here, $\psi(x)$ is the DiGamma function.
\end{widetext}
From (\ref{exenergy}), one has the following relation
between the physical excitation energy $\omega$ and $\Lam^h$,
 $(\omega+H^\prime)/T_k= e^{\frac{\pi}{c}\Lambda^h}$ where
as before $H^\prime =(\frac{e}{2\pi})^{1/2}H$ . Note that as
always, $\omega$ is measured from the Fermi-surface Combining this with
the previously stated relation between $B$ and $H$ the DOS
becomes
\begin{widetext}
\begin{equation}
N(\omega)= \frac{1}{2\pi^2} \lba
\frac{1}{\omega + H^\prime} 
{\rm Re \,}[ \beta (S+ i\frac{1}{\pi}\log{((\omega+H^\prime)/T_k)})] 
+ \frac{H}{2\pi (\omega +H^\prime)^2} {\rm Re \,}[ \beta ( S+ i
\frac{1}{\pi}\log{(H^\prime/T_k)})] \rba
\end{equation}
\end{widetext}


\vfill \eject


\begin{thebibliography}{99}
\bibitem{footnote} Note that the assumption of Ref.~\onlinecite{JanZar}, that the single particle 
scattering cross section vanishes in the $\omega \to 0$ limit is incorrect. The intereference 
of the unscattered electron waves with  the elastically scattered portion of their wave 
function yields a complete cancellation for out-going single particle states.
\bibitem{footnote3} Alternatively, one can think about adding an electron
to the system and seeing the effect this has on the density of solutions.
From simple counting arguments, it can be shown that adding an electron
corresponds to creating a hole $\Lambda^h$ in the density. Hence, at low
energies, the electron can be identified with a spinon.
\end{thebibliography}
\end{document}